\documentclass[aps,reprint,prl,superscriptaddress,showpacs,floats,floatfix]{revtex4-2}
\usepackage{morefloats}
\usepackage{mathptmx}
\usepackage{indentfirst}
\usepackage{amsmath,graphicx,dcolumn}
\usepackage[colorlinks,linkcolor=red, citecolor=blue]{hyperref}
\usepackage[usenames]{color}
\usepackage{datetime}
\usepackage{textcomp}
\usepackage{ulem}
\usepackage{float}

\newcommand{\blue}{\textcolor{blue}}

\newcommand*{\LNO}{La$_4$Ni$_3$O$_{10}$}
\newcommand*{\mjcm}{$\mu$J/cm$^2$}
\begin{document}
	
\title{Nonthermal melting and density wave instability coupled to the lattice in {\LNO}}

\author{Chen Zhang}
\thanks{These authors contributed equally to this work}
\affiliation{School of Physics, Central South University, Changsha 410012, Hunan, China}

\author{Lixing Chen}
\thanks{These authors contributed equally to this work}
\affiliation{State Key Laboratory of Surface Physics and Department of Physics, Fudan University, Shanghai, China}
\affiliation{Shanghai Research Center for Quantum Sciences, Shanghai, China}

\author{Qi-Yi Wu}
\affiliation{School of Physics, Central South University, Changsha 410012, Hunan, China}

\author{Congcong Le}
\affiliation{Hefei National Laboratory, and New Cornerstone Science Laboratory, Hefei, Anhui 230088, China}

\author{Xianxin Wu}
\affiliation{Institute of Theoretical Physics, Chinese Academy of Sciences, Beijing 100190, China}

\author{Hao Liu}
\affiliation{School of Physics, Central South University, Changsha 410012, Hunan, China}

\author{Bo Chen}
\affiliation{School of Physics, Central South University, Changsha 410012, Hunan, China}

\author{Ying Zhou}
\affiliation{School of Physics, Central South University, Changsha 410012, Hunan, China}

\author{Zhong-Tuo Fu}
\affiliation{School of Physics, Central South University, Changsha 410012, Hunan, China}

\author{Chun-Hui Lv}
\affiliation{School of Physics, Central South University, Changsha 410012, Hunan, China}

\author{Zi-Jie Xu}
\affiliation{School of Physics, Central South University, Changsha 410012, Hunan, China}

\author{Hai-Long Deng}
\affiliation{School of Physics, Central South University, Changsha 410012, Hunan, China}

\author{Enkang Zhang}
\affiliation{State Key Laboratory of Surface Physics and Department of Physics, Fudan University, Shanghai, China}
\affiliation{Shanghai Research Center for Quantum Sciences, Shanghai, China}

\author{Yinghao Zhu}
\affiliation{State Key Laboratory of Surface Physics and Department of Physics, Fudan University, Shanghai, China}
\affiliation{Shanghai Research Center for Quantum Sciences, Shanghai, China}

\author{H. Y. Liu}
\affiliation{Beijing Academy of Quantum Information Sciences, Beijing 100085, China}

\author{Yu-Xia Duan}
\affiliation{School of Physics, Central South University, Changsha 410012, Hunan, China}

\author{Jun Zhao}
\email{Corresponding author: zhaoj@fudan.edu.cn}
\affiliation{State Key Laboratory of Surface Physics and Department of Physics, Fudan University, Shanghai, China}
\affiliation{Shanghai Research Center for Quantum Sciences, Shanghai, China}

\author{Jian-Qiao Meng}
\email{Corresponding author: jqmeng@csu.edu.cn}
\affiliation{School of Physics, Central South University, Changsha 410012, Hunan, China}

\date{\today}
	
\begin{abstract}

The recent discovery of high-temperature superconductivity in pressurized nickelates has renewed interest in the broken-symmetry states of their ambient-pressure parent phases, where a density-wave (DW) order emerges and competes with superconductivity, but its microscopic origin remains unresolved. Using ultrafast optical spectroscopy, we track quasiparticle relaxation dynamics across the DW transition at $T_{\rm DW} \approx$ 136 K in trilayer nickelate {\LNO} single crystals, revealing the opening of an energy gap of $\sim$52 meV. Multiple coherent phonons, including $A_g$ modes near 3.88, 5.28, and 2.09 THz, display pronounced mode-selective anomalies across the transition, indicating that the DW is strongly coupled to lattice degrees of freedom and suggesting an important role of electron-phonon coupling. At higher excitation densities, the DW is nonthermally suppressed, producing a temperature-fluence phase diagram that parallels pressure-tuned behavior. These results establish the DW in {\LNO} as a lattice-entangled instability involving multiple phonon modes, and highlight ultrafast optical excitation as a nonequilibrium tuning parameter for suppressing density-wave order in nickelates.
 
\end{abstract}
	
\maketitle

The interplay between superconductivity (SC) and density wave (DW) order is a defining yet unresolved theme in correlated electron systems. In the Ruddlesden-Popper nickelates La$_{n+1}$Ni$_n$O$_{3n+1}$, superconductivity emerges when a native DW order is suppressed under hydrostatic pressure, suggesting that the DW quantum critical point plays a central role in their phase diagram \cite{HSun2023, YZhu2024, JHou2023CPL, YNZhang2024, NNWang2024, QLi2024, MZhang2025PRX, JLi2024}. This competition mirrors the cuprates and iron pnictides, where charge and spin orders are closely entwined with superconductivity \cite{BKeimer2015, PCDai2015}. Notably, tetragonal {\LNO} fails to superconduct even under pressures up to 160 GPa in the absence of a DW instability \cite{MShi2025}, underscoring the intimate link between superconductivity and density wave physics.

The bilayer La$_{3}$Ni$_{2}$O$_{7}$  has been the focus of extensive study, yet sample quality limitations have led to conflicting reports on its DW properties \cite{XChen2024, KChen2024, RKhasanov2025, DZhao2025, ZLiu2023}. In contrast, the trilayer {\LNO} displays high crystalline quality and undergoes a robust spin-charge density wave transition at $T_{\rm DW} \approx$ 136 K  \cite{YZhu2024, JZhang2020}, making it an ideal platform for investigating the ordered state. The central ambiguity, however, is the DW's microscopic origin, with conflicting evidence pointing to either single $d_{z^2}$ or hybridized $d_{x^2-y^2}/d_{z^2}$ orbital physics \cite{MKakoi2024, SDeswal2025, ASuthar2025, MLi2025, XDu2024NP, YZhang2024PRL, SXu2025PRB, SXu2025NC, HXLi2017, YDLi2025}. Clarifying this orbital character is crucial for understanding the mechanism of DW formation. Resolving this requires a probe sensitive to both the electronic gap dynamics and the associated lattice symmetry breaking.

Ultrafast optical spectroscopy provides a powerful complementary approach to address this issue. Its femtosecond resolution allows direct observation of quasiparticle relaxation across an energy gap, while coherent lattice vibrations launched by photoexcitation reveal phonon modes that couple to symmetry breaking \cite{QYWu2025, ZWang2021, YZZhao2023}. By monitoring the temperature evolution of both quasiparticle and phonon responses, ultrafast spectroscopy provides simultaneous dynamical access to electronic and lattice degrees of freedom and further enables investigation of nonequilibrium control pathways inaccessible to static probes.

\begin{figure}[!b]
\begin{center}
\includegraphics[width=0.8\columnwidth]{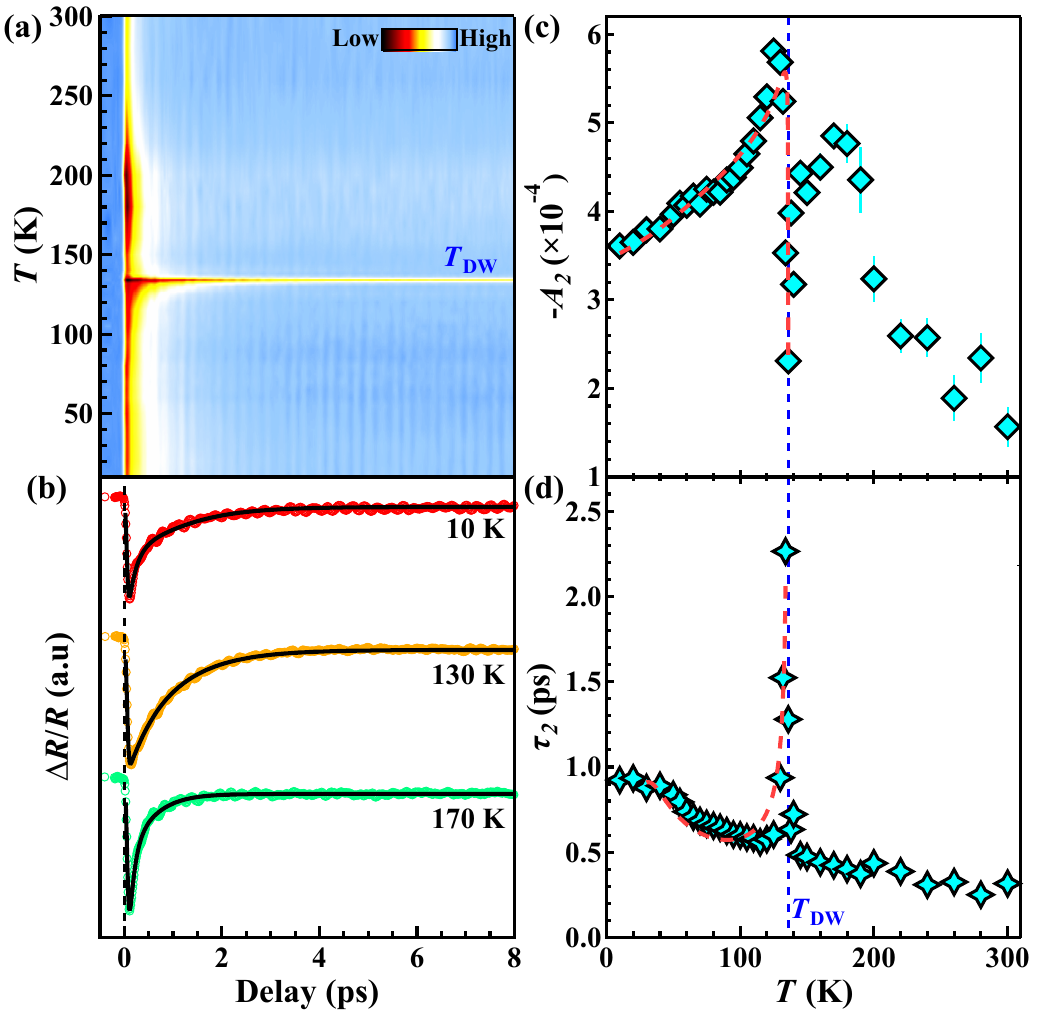}
\end{center}
\setlength{\abovecaptionskip}{-4pt}
\caption{\textbf{Quasiparticle dynamics across the DW transition.} (a) Transient reflectivity $\Delta R(t)/R$ as a function of temperature and time delay, measured at a low pump fluence of $\sim$10.2 {\mjcm}. The sharp change at
$T_{\rm DW}$ marks the phase transition. (b) Representative $\Delta R(t)/R$ traces at different temperatures. Black lines are fits using a biexponential model. (c, d) Temperature dependence of the amplitude ($A_2$) and lifetime ($\tau_2$) of the slow relaxation component. Dashed red curves are fits to the RT model.}
\label{FIG:1}
\end{figure}
\begin{figure*}[hbt]

\begin{center}
\includegraphics[width=1.7\columnwidth]{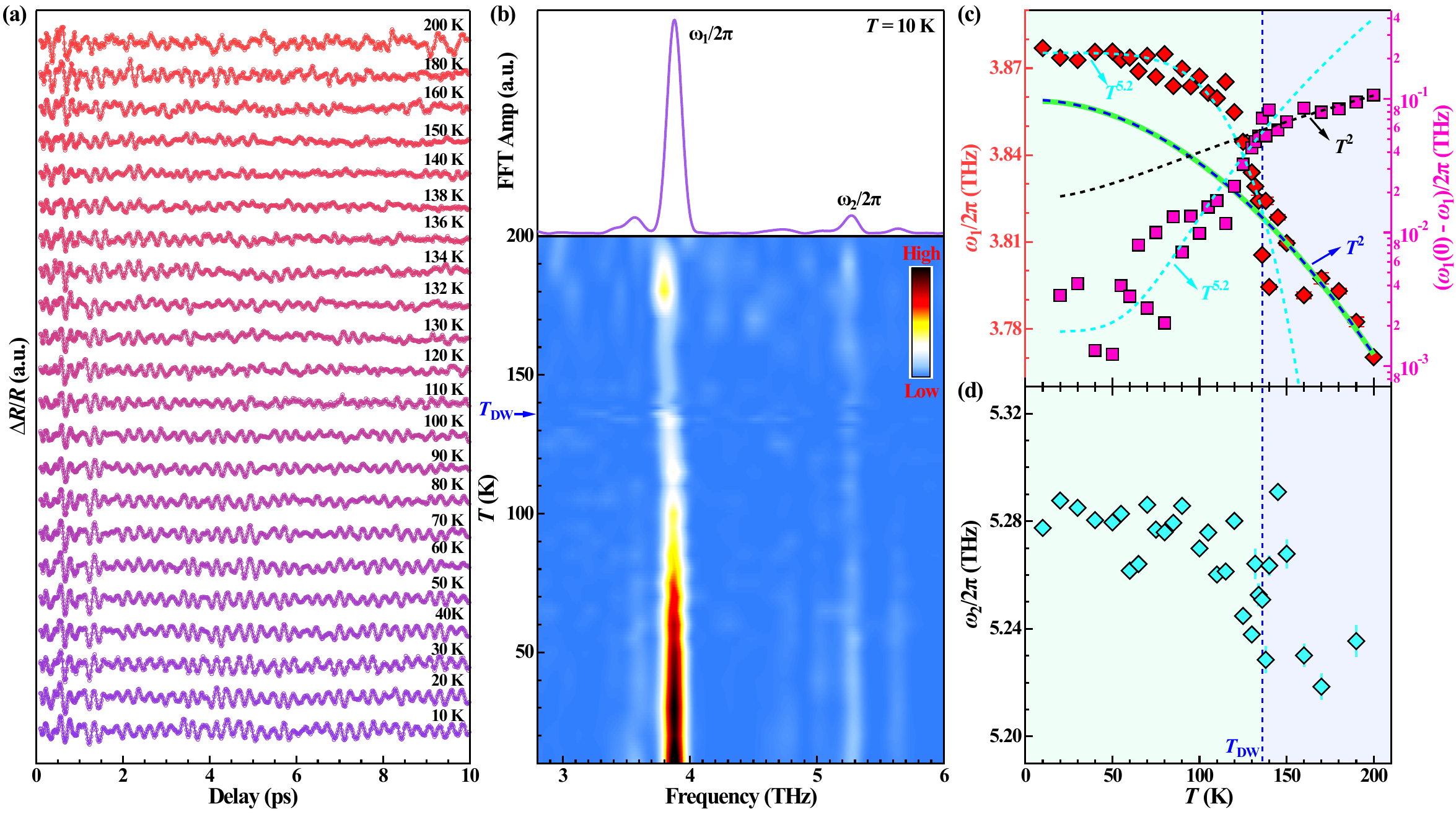}
\end{center}
\setlength{\abovecaptionskip}{-4pt}
\caption{\textbf{Anomalous phonon dynamics at low pump fluence.} (a) Coherent oscillations at select temperatures, isolated by subtracting the quasiparticle background. Traces are offset for clarity. (b) Corresponding FFT map, showing the temperature evolution of the two most prominent phonon modes, $\omega_1$ and $\omega_2$.  (c) Temperature dependence of the $\omega_1$ mode frequency. The solid green line is a fit to a standard anharmonic decay model, from which the data deviate below $T_{\rm DW}$. Dashed curves are power-law fits ($T^2$ above $T_{\rm DW}$, $T^{5.2}$ below), and the right axis shows the data on a log-log scale to highlight the change in slope at the transition. (d) Temperature dependence of the $\omega_2$ mode frequency.}
\label{FIG:2}
\end{figure*}

Here we use ultrafast optical spectroscopy to probe {\LNO}. We report three main findings: (i) Quasiparticle relaxation dynamics reveal the opening of a strong-coupling 52 meV DW gap at $T_{\rm DW}$; (ii) a phonon at 3.88 THz undergoes anomalous renormalization strongly coupled to the DW instability, signaling strong electron-phonon ($e$-ph) coupling and multiorbital involvement; and (iii) intense optical excitation suppresses the DW nonthermally, yielding a temperature-fluence phase diagram analogous to pressure tuning. Together, these results provide insight into the microscopic nature of the DW in {\LNO} and demonstrate light as an effective nonequilibrium tuning parameter for the density wave state in nickelates.

High-quality single crystals of {\LNO} were grown using a high-pressure optical floating-zone furnace (HKZ, SciDre GmbH; 5~kW xenon lamp) under 18-22~bars O$_2$ \cite{YZhu2024}. Transient differential reflectivity $\Delta R(t)/R$ was measured using a pump-probe setup on a 1-MHz Yb-based femtosecond oscillator. The laser output was frequency-doubled and converted by an optical parametric amplifier to provide pump and probe pulses with a central wavelength of 800 nm (1.55 eV) \cite{CZhang2022, QYWu2026}. The overall temporal resolution of the setup is approximately 60 fs, as characterized previously \cite{CZhang2022}, sufficient to resolve the subpicosecond dynamics discussed below. To improve the signal-to-noise ratio, the pump and probe beams are linearly cross-polarized, and a polarizer is placed in front of the detector to suppress scattered pump light. Temperature-dependent weak-fluence measurements were performed on sample S1, while other measurements used sample S2 from the same batch.

Figure \blue{1(a)} maps the transient reflectivity $\Delta R(t)/R$ across the DW transition at a low pump fluence of $\sim$10.2 {\mjcm}. Each trace [Fig. \blue{1(b)}] consists a nonoscillatory quasiparticle (QP) relaxation alongside coherent oscillations. The QP dynamics are well described by a biexponential decay (see Sec. \blue{I} of the Supplemental Material \cite{SUPPM}), indicating two distinct relaxation channels. The fast component ($\tau_1$) is attributed to rapid carrier thermalization and scattering processes, whereas the slower component ($\tau_2$) is associated with quasiparticle recombination across the DW gap. The latter therefore provides a direct probe of the ordered state.

As shown in Figs. \blue{1(c)} and \blue{1(d)}, the amplitude $A_2$ and lifetime $\tau_2$ exhibit sharp anomalies at $T_{\rm DW} \approx$ 136 K. Most notably, $\tau_2$ exhibits pronounced slowing down near $T_{\rm DW}$, which is consistent with critical dynamics commonly observed near continuous density wave transitions. This dynamic signature confirms the opening of an energy gap \cite{QYWu2025, ZWang2021}. Within the Rothwarf-Taylor (RT) framework \cite{ARothwarf1967, VKabanov1999}, the DW gap is primarily extracted from the temperature dependence of the slow-component amplitude $A_2(T)$, while $\tau_2(T)$ serves as a consistency check of the phonon-bottleneck relaxation dynamics. This analysis yields a zero-temperature gap of 2$\Delta_{\rm DW}(0) \approx$ 52 meV (see Sec. \blue{II} of the Supplemental Material \cite{SUPPM}), consistent with some other spectroscopic studies \cite{SXu2025PRB, SXu2025NC}.

The ratio 2$\Delta_{\rm DW}(0)/k_BT_{\rm DW}\approx$ 4.4 exceeds the weak-coupling BCS value, identifying {\LNO} as a strong-coupling system. This ratio, however, is substantially smaller than that found in bilayer La$_3$Ni$_2$O$_7$ ($\sim$11) \cite{QYWu2025B}, pointing to fundamental differences between trilayer and bilayer nickelates in how density wave order is stabilized.

\begin{figure*}[ht]
\begin{center}
\includegraphics[width=1.5\columnwidth]{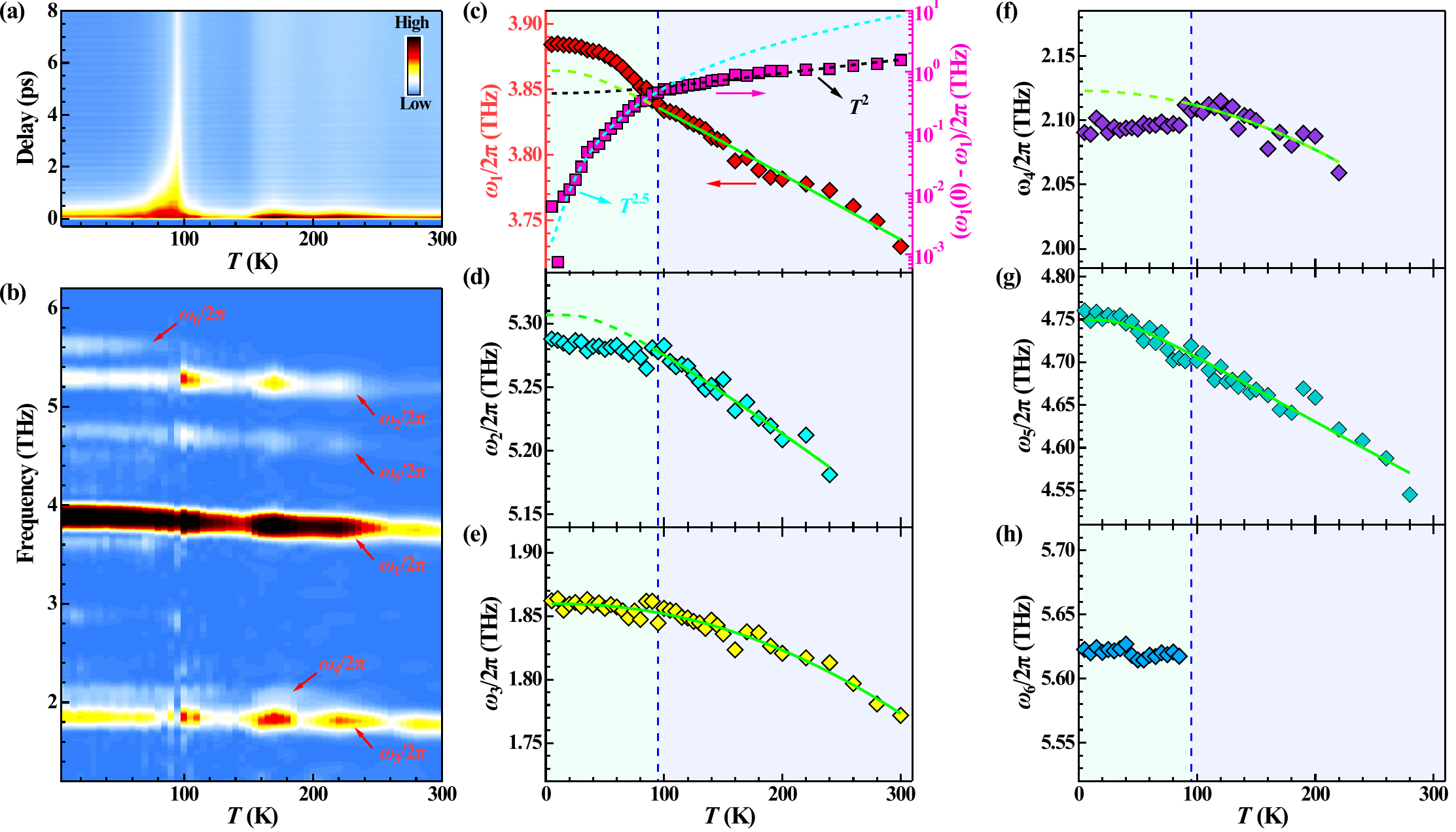}
\end{center}
\setlength{\abovecaptionskip}{-4pt}
\caption{ \textbf{Full phonon spectrum and mode-selective coupling at high pump fluence.} (a) Raw $\Delta R(t)/R$ map measured at a high pump fluence of $\sim $130 {\mjcm}, which suppresses the transition to $T_{\rm DW} \approx$ 95 K. (b) Corresponding FFT map revealing six distinct coherent phonon modes ( $\omega_1$ to  $\omega_6$). (c-h) Temperature dependence of the frequency for each of the six modes. Vertical dashed lines indicate the suppressed $T_{\rm DW}$. Solid green lines are fits to the anharmonic decay model, highlighting the anomalous behavior of modes $\omega_1$ and $\omega_4$ below the transition.}
\label{FIG:3}
\end{figure*}

Having characterized the electronic gap, we now analyze the coherent lattice oscillations to examine $e$-ph coupling. Figure \blue{2(a)} presents the time-domain traces after subtracting the QP background, while Fig. \blue{2(b)} shows the corresponding fast Fourier transform (FFT) spectra. Two prominent $A_g$ symmetry phonons are resolved (see Sec. \blue{III} of the Supplemental Material \cite{SUPPM}): $\omega_1 \approx$ 3.88 THz (129.4 cm$^{-1}$) and $\omega_2 \approx$ 5.28 THz (176.1 cm$^{-1}$) at 10 K, in agreement with Raman measurements \cite{SDeswal2025, ASuthar2025}. Both modes soften with increasing temperature, consistent with previous ultrafast measurements on {\LNO} \cite{YDLi2025}. A more detailed analysis, however, reveals that the $\omega_1$ mode deviates from conventional anharmonic behavior near $T_{\rm DW}$. Above the transition, $\omega_1$ follows standard anharmonic phonon decay \cite{MBalkanski1983, JMenendez1984}, whereas below $T_{\rm DW}$ it deviates sharply, showing a much stronger temperature dependence [Fig. \blue{2(c)}]. A power-law analysis [$\omega_1(T) = \omega_1(0) - \alpha T^n$], with $\omega_1(0)$ the extrapolated zero-temperature frequency, quantifies this change \cite{JHasaien2025}: the exponent $n$ switches from $n$ = 2 (anharmonic theory) above $T_{\rm DW}$ to $n \approx$ 5.2 below it, which is highlighted in the log-log plot shown on the right axis of Fig. \blue{2(c)}. This abrupt renormalization indicates strong coupling between the $\omega_1$ phonon and the density wave instability, although the mode does not exhibit the behavior expected of a conventional DW amplitude mode. Besides these two modes, a weak shoulderlike feature is visible near 3.55 THz on the low-frequency side of the dominant $\omega_1$ mode. Its apparent intensity is comparable to that of the weak $\omega_2$ mode in the low-fluence FFT map [Fig. \blue{2(b)}], but it becomes much weaker than $\omega_2$ under high-fluence excitation [Fig. \blue{3(b)}]. Because this feature is strongly affected by the tail of the intense $\omega_1$ peak and cannot be reliably tracked as a function of temperature, we do not assign it as an independent coherent phonon mode in the present analysis.

We also note that the oscillation amplitude of the $\omega_1$ phonon is strongly suppressed near $T_{\rm DW}$ but partially recovers upon further warming above the transition. While the microscopic origin of this behavior remains unclear, similar behavior has been reported in Raman measurements on {\LNO} \cite{SDeswal2025}. Possible contributing factors include temperature-dependent changes in electronic screening and $e$-ph coupling across the transition, which may modify the optical activity of this phonon mode even in the high-temperature phase.

To uncover weaker lattice modes, we increased the pump fluence to 130 {\mjcm} (Fig. \blue{3}).Under these conditions, the transition temperature is suppressed to $T_{\rm DW} \approx$ 95 K [Fig. \blue{3(a)}]. Within the RT analysis framework, the characteristic gap scale inferred from the relaxation amplitude exhibits much weaker fluence dependence than $T_{\rm DW}$, yielding 2$\Delta_{\rm DW}(0) \approx$ 55 meV  (see Sec. \blue{II} of the Supplemental Material \cite{SUPPM}) and an enhanced apparent ratio 2$\Delta_{\rm DW}(0)/k_BT_{\rm DW} \approx$ 6.6, compared with $\sim$4.4 at low fluence. This behavior contrasts with hydrostatic pressure, where both $T_{\rm DW}$ and the gap are monotonically reduced \cite{SXu2025NC}, and with ultrafast measurements on bilayer La$_3$Ni$_2$O$_7$, where the gap diminishes under increasing fluence, leading to a reduced ratio \cite{QYWu2025B}. Such differences likely reflect the distinct nature of equilibrium versus nonequilibrium perturbations, as well as material-dependent variations in dimensionality and electron-lattice coupling. In particular, the pronounced lattice entanglement observed here suggests that the DW state in {\LNO} may be more strongly lattice stabilized, contributing to its unusual robustness against photodoping.

\begin{figure*}[ht]
\begin{center}
\includegraphics[width=1.3\columnwidth]{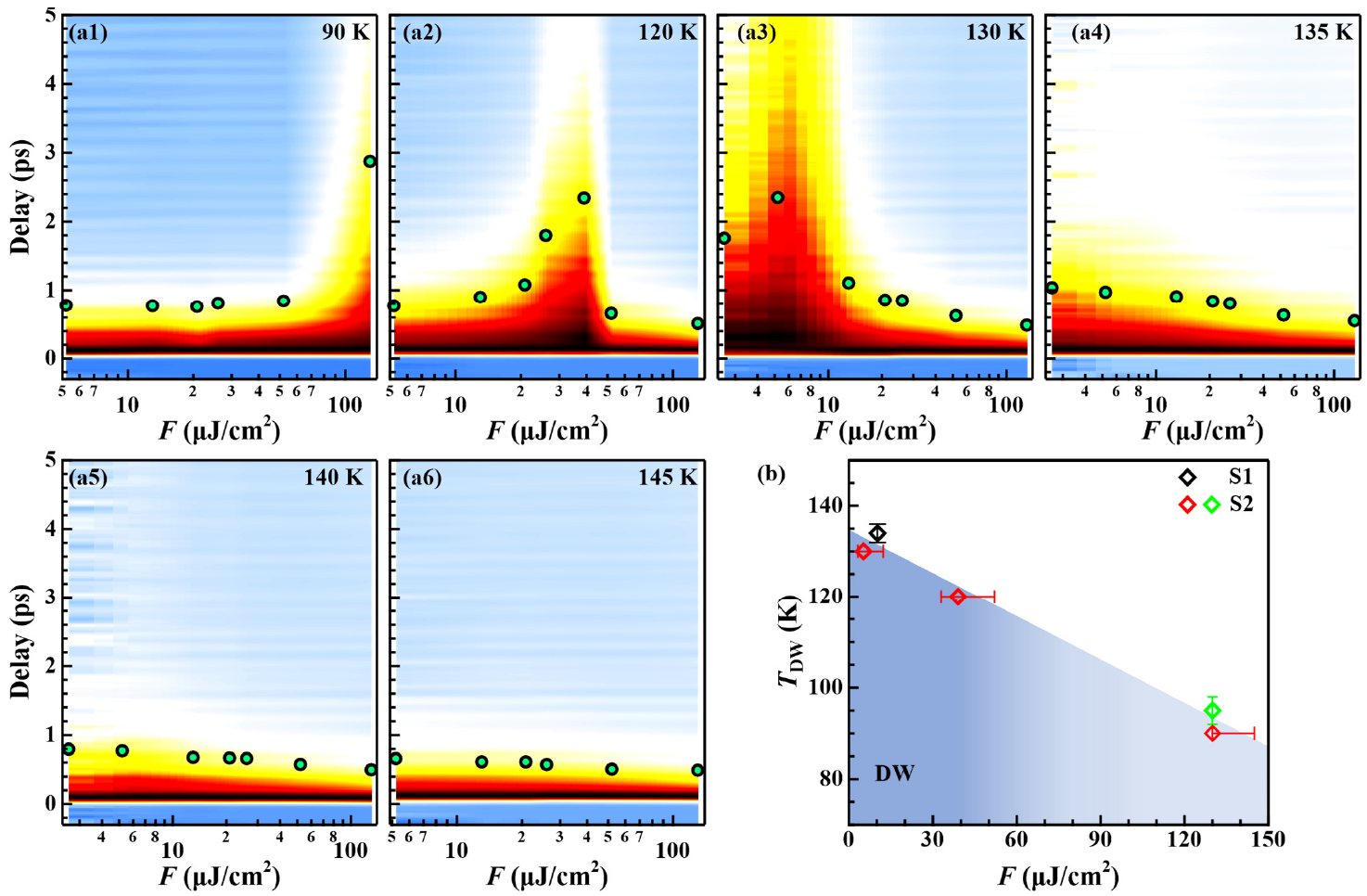}
\end{center}
\setlength{\abovecaptionskip}{-4pt}
\caption{\textbf{Non-thermal melting of the DW order.} (a1)-(a6) Normalized $\Delta R(t)/R$ maps as a function of pump fluence at several fixed temperatures. The black circles trace the slow relaxation time $\tau_2$, which peaks at the critical fluence $F_C$ required to melt the DW order. (b) The resulting Temperature-Fluence phase diagram for {\LNO}. The phase boundary is determined by the $F_C(T)$ values, extracted from the peaks in $\tau_2$ shown in (a), together with data from Figs. \blue{1} (black  symbol) and \blue{3} (green symbol).}
\label{FIG:4}
\end{figure*}

At high fluence, four additional phonon ($\omega_3–\omega_6$) resolved [Fig. \blue{3(b)}] (see Sec. \blue{III} of the Supplemental Material \cite{SUPPM}), revealing a richer picture of mode-selective coupling [Figs. \blue{3(c)–3(h)}]. Notably, multiple coherent phonon modes exhibit a common dip in amplitude near $\sim$145 K, slightly above the equilibrium $T_{\rm DW}\approx$ 136 K. Because this feature is absent at low fluence and appears simultaneously across several modes, it likely reflects a high-fluence-induced modification of the nonequilibrium electronic response near the DW transition rather than mode-specific lattice anomalies. Despite this additional complexity, the $\omega_1$ mode retains its anomalous renormalization at the reduced $T_{\rm DW}$, transitioning from $T^2$ behavior above the transition to $T^{2.5}$ scaling within the ordered phase [Fig. \blue{3(c)}]. While the precise exponent depends somewhat on excitation density, the robust crossover at $T_{\rm DW}$ and the consistent $T^2$ scaling above it support a close association between the $\omega_1$ phonon and the DW instability. Its eigenvector involves apical oxygen motion, Ni-O plane buckling, and La displacements \cite{ASuthar2025, YDLi2025}, indicating that the lattice response associated with the DW instability involves multiple structural degrees of freedom. Combined with prior angle-resolved photoemission spectroscopy (ARPES) and theoretical studies suggesting hybridized orbital character of the DW-gapped electronic states \cite{XDu2024NP}, this behavior is consistent with the relevance of multiorbital physics in the DW instability of {\LNO}.

Importantly, the $\omega_2$ phonon [Fig. \blue{3(d)}] also exhibits anomalous renormalization, but in the opposite direction: Instead of hardening as $\omega_1$, $\omega_2$ softens below $T_{\rm DW}$. The contrasting behavior of these two $A_g$ modes indicates that the density wave order couples to multiple phonons in distinct ways, highlighting the complex, mode-dependent nature of $e$-ph interactions in {\LNO}.

Another key anomaly appears in the $\omega_4$ mode ($\approx$2.09 THz) [Fig.  \blue{3(f)}], which softens significantly below $T_{\rm DW}$. This behavior resembles phonon renormalization observed near charge-ordering transitions in several cuprate systems \cite{BLoret2019, DOh2021, HYHuang2021}, where gap opening can modify $e$-ph self-energy effects and lead to anomalous phonon softening \cite{DOh2021}. Such similarity suggests that related $e$-ph coupling effects may be operative in the DW state of {\LNO}. By contrast, the $\omega_3$ ($\approx$1.86 THz) and $\omega_5$ ($\approx$4.76 THz) modes [Figs. \blue{3(e)} and \blue{3(g)}] follow ordinary anharmonic decay, serving as internal controls that confirm the selectivity of the anomalous coupling.

Finally, the $\omega_6$ mode ($\approx$5.62 THz) appears only below $T_{\rm DW}$ [Fig. \blue{3(h)}]. Its sudden onset rules out a conventional amplitude mode and suggests alternative origins, such as reduced metallic screening due to partial Fermi-surface gapping \cite{ASuthar2025} or zone folding from symmetry lowering \cite{XFTang2022}. Its weak intensity in our data prevents a conclusive assignment, motivating further experiments.

Beyond serving as a probe, the pump laser also acts as a tuning parameter for the ordered phase \cite{ZWang2021, SDuan2021, HLiu2025}. Figures \blue{4(a1)}–\blue{4(a6)} show how the QP relaxation evolves with fluence at fixed temperatures. Below $T_{\rm DW}$, the slow component $\tau_2$ increases with fluence, peaks at a critical fluence $F_C$, and then collapses, signaling complete melting of the density wave condensate. The resulting temperature-fluence phase diagram [Fig. \blue{4(b)}] shows that $F_C$ grows as the system is cooled, increasing from $\sim$5.2 {\mjcm} at 130 K to over 130 {\mjcm} at 90 K. This behavior resembles pressure-tuned suppression of the DW \cite{YHMeng2024, SXu2025NC}, but with a crucial difference: under pressure, both $T_{\rm DW}$ and the gap decrease together \cite{YHMeng2024, SXu2025NC}, while in La$_3$Ni$_2$O$_7$ optical excitation suppresses them simultaneously \cite{QYWu2025B}. In {\LNO}, however, light reduces $T_{\rm DW}$ while preserving the gap, yielding an enhanced ratio. Even at the highest fluence, the steady-state temperature rise is estimated to be only $\sim$8 K (see Sec. \blue{IV} of Supplemental Material \cite{SUPPM}), substantially smaller than the observed $\sim$40 K suppression of $T_{\rm DW}$. While thermal effects may contribute partially, the discrepancy indicates that simple laser heating alone cannot account for the observed behavior. These results therefore indicate a substantial nonequilibrium contribution to the photoinduced suppression of the DW order, providing a distinctly nonequilibrium pathway to tune ordered states in nickelates.

 Experimental data indicate an intrinsic coupling between the DW order and lattice degrees of freedom in trilayer nickelates. The DW state comprises intertwined spin density wave (SDW) and charge density wave (CDW): The SDW component is dominantly localized on the outer NiO$_2$ layers, whereas the CDW modulates all three layers\cite{JZhang2020}. The key difference between the $\omega_1$ and $\omega_2$ phonon modes lies in the distinct motion of the inner-layer oxygen atoms\cite{SUPPM}. Consequently, the distinct temperature dependences of $\omega_1$ and $\omega_2$ phonon frequencies below $T_{\rm DW}$ point to a selective coupling of the lattice to the CDW component of the DW state. This interpretation is consistent with the observed isotope dependence of T$_{\text{DW}}$ in La$_4$Ni$_3$O$_{10}$ \cite{Khasanov2025}, which directly implicates phonons. Taken together, these observations suggest that $e$-ph coupling likely play a significant role in stabilizing the DW order, and may also contribute superconductivity once the DW order is suppressed.

In summary, ultrafast optical spectroscopy of {\LNO} reveals the opening of a strong-coupling density wave gap, anomalous renormalization of a lattice mode strongly coupled to the DW instability, and nonthermal optical melting of the ordered state. The contrasting behavior of phonon modes highlights mode-dependent electron-phonon coupling and the relevance of multiorbital physics in this material. These findings establish the DW in {\LNO} as a lattice-entangled instability and demonstrate that ultrafast optical excitation provides a powerful nonequilibrium pathway to suppress and tune density wave order in nickelates.

This work was supported by the National Natural Science Foundation of China (Grant No. 92265101), the National Key Research and Development Program of China (Grant No. 2022YFA1604204), the Open Project of Beijing National Laboratory for Condensed Matter Physics (Grant No. 2024BNLCMPKF001), and the Science and Technology Innovation Program of Hunan province (2022RC3068). The work at Fudan University was supported by the Key Program of the National Natural Science Foundation of China (Grant No. 12234006), the National Key Research and Development Program of China (Grant No. 2022YFA1403202), Quantum Science and Technology-National Science and Technology Major Project (Grant No. 2024ZD0300103), the Shanghai Municipal Science and Technology Major Project (Grant No. 2019SHZDZX01), and the Large Scientific Facility Open Subject of Songshan Lake Laboratory (Grant No. KFKT2022A03).


\begin{thebibliography}{100}

\bibitem{HSun2023}
H. Sun, M. Huo, X. Hu, J. Li, Z. Liu, Y. Han, L. Tang, Z. Mao, P. Yang, B. Wang, $et$ $al$., Signatures of superconductivity near 80 K in a nickelate under high pressure, \href{https://www.nature.com/articles/s41586-023-06408-7}{\textcolor{blue}{Nature \textbf{621}, 493 (2023)}}.

\bibitem{YZhu2024}
Y. Zhu, D. Peng, E. Zhang, B. Pan, X. Chen, L. Chen, H. Ren, F. Liu, Y. Hao, N. Li $et$ $al$., Superconductivity in pressurized trilayer La$_4$Ni$_3$O$_{10{-\delta}}$ single crystals, \href{https://www.nature.com/articles/s41586-024-07553-3}{\textcolor{blue}{Nature \textbf{631}, 531 (2024)}}.

\bibitem{JHou2023CPL}
J. Hou, P. T. Yang, Z. Y. Liu, J. Y. Li, P. F. Shan, L. Ma, G. Wang, N. N. Wang, H. Z. Guo, J.-P. Sun $et$ $al$., Emergence of High-Temperature Superconducting Phase in Pressurized La$_3$Ni$_2$O$_7$ Crystals, \href{https://cpl.iphy.ac.cn/article/10.1088/0256-307X/40/11/117302}{\textcolor{blue}{Chin. Phys. Lett. \textbf{40}, 117302 (2023)}}.

\bibitem{YNZhang2024}
Y. Zhang, D. Su, Y. Huang, Z. Shan, H. Sun, M. Huo, K. Ye, J. Zhang, Z. Yang, Y. Xu $et$ $al$., High-temperature superconductivity with zero resistance and strange-metal behaviour in La$_3$Ni$_2$O$_{7{-\delta}}$, \href{https://www.nature.com/articles/s41567-024-02515-y}{\textcolor{blue}{Nat. Phys. \textbf{20}, 1269 (2024)}}.

\bibitem{NNWang2024}
N. Wang, G. Wang, X. Shen, J. Hou, J. Luo, X. Ma, H. Yang, L. Shi, J. Dou, J. Feng $et$ $al$., Bulk high-temperature superconductivity in pressurized tetragonal La$_2$PrNi$_2$O$_7$, \href{https://www.nature.com/articles/s41586-024-07996-8}{\textcolor{blue}{Nature \textbf{634}, 579 (2024)}}.

\bibitem{QLi2024}
Q. Li, Y. J. Zhang, Z. N. Xiang, Y. Zhang, X. Zhu, and H. H. Wen, Signature of Superconductivity in Pressurized La$_4$Ni$_3$O$_{10}$, \href{https://cpl.iphy.ac.cn/article/10.1088/0256-307X/41/1/017401}{\textcolor{blue}{Chin. Phys. Lett. \textbf{41}, 017401 (2024)}}.

\bibitem{MZhang2025PRX}
M. Zhang, C. Pei, D. Peng, X. Du, W. Hu, Y. Cao, Q. Wang, J. Wu, Y. Li, H. Liu $et$ $al$., Superconductivity in Trilayer Nickelate La$_4$Ni$_3$O$_{10}$ under Pressure, \href{https://doi.org/10.1103/PhysRevX.15.021005}{\textcolor{blue}{Phys. Rev. X \textbf{15}, 021005 (2025)}}.

\bibitem{JLi2024}
J. Li, C. Q. Chen, C. Huang, Y. Han, M. Huo, X. Huang, P. Ma, Z. Qiu, J. Chen, X. Hu $et$ $al$., Structural transition, electric transport, and electronic structures in the compressed trilayer nickelate La$_4$Ni$_3$O$_{10}$, \href{https://link.springer.com/article/10.1007/s11433-023-2329-x}{\textcolor{blue}{Sci. China-Phys. Mech. Astron. \textbf{67}, 117403 (2024)}}.

\bibitem{BKeimer2015}
B. Keimer, S. A. Kivelson, M. R. Norman, S. Uchida, and J. Zaanen, From quantum matter to high-temperature superconductivity in copper oxides, \href{https://www.nature.com/articles/nature14165}{\textcolor{blue}{Nature \textbf{518}, 179 (2015)}}.

\bibitem{PCDai2015}
P. Dai, Antiferromagnetic order and spin dynamics in iron-based superconductors, \href{https://doi.org/10.1103/RevModPhys.87.855}{\textcolor{blue}{Rev. Mod. Phys. \textbf{87}, 855 (2015)}}.

\bibitem{MShi2025}
M. Shi, Y. Li, Y. Wang, D. Peng, S. Yang, H. Li, K. Fan, K. Jiang, J. He, Q. Zeng $et$ $al$., Absence of superconductivity and density-wave transition in ambient-pressure tetragonal La$_4$Ni$_3$O$_{10}$, \href{https://www.nature.com/articles/s41467-025-57264-0}{\textcolor{blue}{Nat. Commun. \textbf{16}, 2887 (2025)}}.

\bibitem{XChen2024}
X. Chen, J. Choi, Z. Jiang, J. Mei, K. Jiang, J. Li, S. Agrestini, M. Garcia-Fernandez, H. Sun, X. Huang $et$ $al$., Electronic and magnetic excitations in La$_3$Ni$_2$O$_7$, \href{https://www.nature.com/articles/s41467-024-53863-5}{\textcolor{blue}{Nat. commun. \textbf{15}, 9597 (2024)}}.

\bibitem{KChen2024}
K. Chen, X. Liu, J. Jiao, M. Zou, C. Jiang, X. Li, Y. Luo, Q. Wu, N. Zhang, Y. Guo $et$ $al$., Evidence of Spin Density Waves in La$_3$Ni$_2$O$_{7{-\delta}}$, \href{https://doi.org/10.1103/PhysRevLett.132.256503}{\textcolor{blue}{Phys. Rev. Lett. \textbf{132}, 256503 (2024)}}.

\bibitem{RKhasanov2025}
R. Khasanov, T. J. Hicken, D. J. Gawryluk, V. Sazgari, I. Plokhikh, L. P. Sorel, M. Bartkowiak, S. Botzel, F. Lechermann, I. M. Eremin $et$ $al$., Pressure-enhanced splitting of density wave transitions in La$_3$Ni$_2$O$_{7{-\delta}}$, \href{https://www.nature.com/articles/s41567-024-02754-z}{\textcolor{blue}{Nat. Phys. \textbf{21}, 430 (2025)}}.

\bibitem{DZhao2025}
D. Zhao, Y. Zhou, M. Huo, Y. Wang, L. Nie, Y. Yang, J. Ying, M. Wang, T. Wu, and X. Chen, Pressure-enhanced spin-density-wave transition in double-layer nickelate La$_3$Ni$_2$O$_{7{-\delta}}$, \href{https://doi.org/10.1016/j.scib.2025.02.019}{\textcolor{blue}{Sci. Bull. \textbf{70}, 1239 (2025)}}.

\bibitem{ZLiu2023}
Z. Liu, H. Sun, M. Huo, X. Ma, Y. Ji, E. Yi, L. Li, H. Liu, J. Yu, Z. Zhang $et$ $al$., Evidence for charge and spin density waves in single crystals of La$_3$Ni$_2$O$_7$ and La$_3$Ni$_2$O$_6$, \href{https://link.springer.com/article/10.1007/s11433-022-1962-4}{\textcolor{blue}{Sci. China-Phys. Mech. Astron. \textbf{66}, 217411 (2023)}}.

\bibitem{JZhang2020}
J. Zhang, D. Phelan, A. Botana, Y.-S. Chen, H. Zheng, M. Krogstad, S. G. Wang, Y. Qiu, J. Rodriguez-Rivera, R. Osborn $et$ $al$., Intertwined density waves in a metallic nickelate, \href{https://www.nature.com/articles/s41467-020-19836-0}{\textcolor{blue}{Nat. Commun. \textbf{11}, 6003 (2020)}}.

\bibitem{MKakoi2024}
M. Kakoi, T. Oi, Y. Ohshita, M. Yashima, K. Kuroki, T. Kato, H. Takahashi, S. Ishiwata, Y. Adachi, N. Hatada $et$ $al$., Multiband Metallic Ground State in Multilayered Nickelates La$_3$Ni$_2$O$_7$ and La$_4$Ni$_3$O$_{10}$ Probed by $^{139}$La-NMR at Ambient Pressure, \href{https://journals.jps.jp/doi/full/10.7566/JPSJ.93.053702}{\textcolor{blue}{Phys. Soc. Jpn. \textbf{93}, 053702 (2024)}}.

\bibitem{SDeswal2025}
S. Deswal, D. Kumar, D. Rout, S. Singh, and P. Kumar, Dynamics of electron–electron correlation and electron–phonon coupled phase progression in trilayer nickelate La$_4$Ni$_3$O$_{10}$, \href{https://doi.org/10.1063/5.0288265}{\textcolor{blue}{Appl. Phys. Lett. \textbf{127}, 071903 (2025)}}. 

\bibitem{ASuthar2025}
A. Suthar, V. Sundaramurthy, M. Bejas, C. Le, P. Puphal, P. Sosa-Lizama, A. Schulz, J. Nuss, M. Isobe, P. A. van Aken $et$ $al$., Multiorbital character of the density wave instability in La$_4$Ni$_3$O$_{10}$, \href{https://arxiv.org/abs/2508.06440}{\textcolor{blue}{arXiv:2508.06440 (2025)}}.

\bibitem{MLi2025}
M. Li, J. Gong, Y. Zhu, Z. Chen, J. Zhang, E. Zhang, Y. Li, R. Yin, S. Wang, J. Zhao $et$ $al$., Direct visualization of an incommensurate unidirectional charge density wave in La$_4$Ni$_3$O$_{10}$, \href{https://doi.org/10.1103/2p56-xl41}{\textcolor{blue}{Phys. Rev. B \textbf{112}, 045132 (2025)}}. 

\bibitem{XDu2024NP}
X. Du, Y. Li, Y. Cao, C. Pei, M. Zhang, W. Zhao, K. Zhai, R. Xu, Z. Liu, Z. Li $et$ $al$., Correlated Electronic Structure and Density-Wave Gap in Trilayer Nickelate La$_4$Ni$_3$O$_{10}$, \href{https://arxiv.org/abs/2405.19853}{\textcolor{blue}{arXiv:2405.19853 (2024)}}. 

\bibitem{YZhang2024PRL}
Y. Zhang, L. F. Lin, A. Moreo, T. A. Maier, and E. Dagotto, Prediction of $s^{{\pm}-}$Wave Superconductivity Enhanced by Electronic Doping in Trilayer Nickelates La$_4$Ni$_3$O$_{10}$ under Pressure, \href{https://doi.org/10.1103/PhysRevLett.133.136001}{\textcolor{blue}{Phys. Rev. Lett. \textbf{133}, 136001 (2024)}}. 

\bibitem{SXu2025PRB}
S. Xu, C. Q. Chen, M. Huo, D. Hu, H. Wang, Q. Wu, R. Li, D. Wu, M. Wang, D. X. Yao $et$ $al$., Origin of the density wave instability in trilayer nickelate La$_4$Ni$_3$O$_{10}$ revealed by optical and ultrafast spectroscopy, \href{https://doi.org/10.1103/PhysRevB.111.075140}{\textcolor{blue}{Phys. Rev. B \textbf{111}, 075140 (2025)}}. 

\bibitem{SXu2025NC}
S. Xu, H. Wang, M. Huo, D. Hu, Q. Wu, L. Yue, D. Wu, M. Wang, T. Dong, and N. Wang, Collapse of density wave and emergence of superconductivity in pressurized-La$_4$Ni$_3$O$_{10}$ evidenced by ultrafast spectroscopy, \href{https://www.nature.com/articles/s41467-025-62294-9}{\textcolor{blue}{Nat. Commun. \textbf{16}, 7039 (2025)}}. 

\bibitem{HXLi2017}
H. Li, X. Zhou, T. Nummy, J. Zhang, V. Pardo, W. E. Pickett, J. F. Mitchell, and D. S. Dessau, Fermiology and electron dynamics of trilayer nickelate La$_4$Ni$_3$O$_{10}$, \href{https://www.nature.com/articles/s41467-017-00777-0}{\textcolor{blue}{Nat. Commun. \textbf{8}, 704 (2017)}}.

\bibitem{YDLi2025}
Y. Li, Y. Cao, L. Liu, P. Peng, H. Lin, C. Pei, M. Zhang, H. Wu, X. Du, W. Zhao $et$ $al$., Distinct ultrafast dynamics of bilayer and trilayer nickelate superconductors regarding the density-wave-like transitions, \href{https://doi.org/10.1016/j.scib.2024.10.011}{\textcolor{blue}{Sci. Bull. \textbf{70}, 180 (2025)}}. 

\bibitem{QYWu2025}
Q. Y. Wu, C. Zhang, B. Z. Li, H. Liu, J. J. Song, B. Chen, H. Y. Liu, Y. X. Duan, J. He, J. Liu $et$ $al$., Interplay of electron-phonon coupling, pseudogap, and superconductivity in CsCa$_2$Fe$_4$As$_4$F$_2$ studied using ultrafast optical spectroscopy, \href{https://doi.org/10.1103/PhysRevB.111.L081110}{\textcolor{blue}{Phys. Rev. B \textbf{111}, L081110 (2025)}}. 

\bibitem{ZWang2021}
Z. Wang, Q. Wu, Q. Yin, C. Gong, Z. Tu, T. Lin, Q. Liu, L. Shi, S. Zhang, D. Wu $et$ $al$., Unconventional charge density wave and photoinduced lattice symmetry change in the kagome metal CsV$_3$Sb$_5$ probed by time-resolved spectroscopy, \href{https://journals.aps.org/prb/abstract/10.1103/PhysRevB.104.165110}{\textcolor{blue}{Phys. Rev. B \textbf{104}, 165110 (2021)}}. 

\bibitem{YZZhao2023}
Y. Z. Zhao, Q. Y. Wu, C. Zhang, B. Chen, W. Xia, J. J. Song, Y. H. Yuan, H. Liu, F. Y. Wu, X. Q. Ye $et$ $al$., Coupling of optical phonons with Kondo effect and magnetic order in the antiferromagnetic Kondo-lattice compound CeAuSb$_2$, \href{https://doi.org/10.1103/PhysRevB.104.165110}{\textcolor{blue}{Phys. Rev. B \textbf{108}, 075115 (2023)}}. 

\bibitem{CZhang2022}
C. Zhang, Q. Y. Wu, W. S. Hong, H. Liu, S. X. Zhu, J. J. Song, Y. Z. Zhao, F. Y. Wu, Z. T. Liu, S. Y. Liu $et$ $al$., Ultrafast optical spectroscopy evidence of pseudogap and electron-phonon coupling in an iron-based superconductor KCa$_2$Fe$_4$As$_4$F$_2$, \href{https://link.springer.com/article/10.1007/s11433-021-1830-9}{\textcolor{blue}{Sci. China-Phys. Mech. Astron. \textbf{65}, 237411 (2022)}}. 

\bibitem{QYWu2026}
Q. Y. Wu, D. Y. Hu, C. Zhang, H. Liu, B. Chen, Y. Zhou, Z. T. Fu, C. H. Lv, Z. J. Xu, H. L. Deng $et$ $al$., Electronic Nematicity Revealed by Polarized Ultrafast Spectroscopy in Bilayer La$_3$Ni$_2$O$_7$, \href{https://www.sciengine.com/SCPMA/doi/10.1007/s11433-026-3026-3}{\textcolor{blue}{Sci. China-Phys. Mech. Astron. (2026)}}

\bibitem{SUPPM}
See Supplemental Material for additional data of {\LNO}, which includes Refs. \cite{ARothwarf1967, VKabanov1999, JDemsar2006, DRout2020}.

\bibitem{ARothwarf1967}
A.	Rothwarf and B. Taylor, Measurement of Recombination Lifetimes in Superconductors, \href{https://doi.org/10.1103/PhysRevLett.19.27}{\textcolor{blue}{Phys. Rev. Lett. \textbf{19}, 27 (1967)}}. 

\bibitem{VKabanov1999}
V. Kabanov, J. Demsar, B. Podobnik, and D. Mihailovic, Quasiparticle relaxation dynamics in superconductors with different gap structures: Theory and experiments on YBa$_2$Cu$_3$O$_{7{-\delta}}$, \href{https://doi.org/10.1103/PhysRevB.59.1497}{\textcolor{blue}{Phys. Rev. B \textbf{59}, 1497 (1999)}}. 

\bibitem{JDemsar2006}
J. Demsar, J. L. Sarrao, A. J. Taylor, Dynamics of photoexcited quasiparticles in heavy electron compounds, \href{https://iopscience.iop.org/article/10.1088/0953-8984/18/16/R01}{\textcolor{blue}{J. Phys.: Condens. Matter \textbf{18}, R281-R314 (2006)}}.

\bibitem{DRout2020}
D. Rout, S. R. Mudi, M. Hoffmann, S. Spachmann, R. Klingeler, S. Singh, Structural and physical properties of trilayer nickelates $R_4$Ni$_3$O$_{10}$ ($R$=La, Pr, and Nd), \href{https://doi.org/10.1103/PhysRevB.102.195144}{\textcolor{blue}{Phys. Rev. B \textbf{102}, 195144 (2020)}}.

\bibitem{QYWu2025B}
Q. Y. Wu, D. Y. Hu, C. Zhang, M. W. Huo, H. Liu, B. Chen, Y. Zhou, Z. T. Fu $et$ $al$., Ultrafast optical evidence of coexisting density waves in bilayer nickelate La$_3$Ni$_2$O$_7$, \href{https://doi.org/10.1103/j6hl-t2sh}{\textcolor{blue}{Phys. Rev. B \textbf{112}, 235110 (2025)}}.

\bibitem{MBalkanski1983}
M. Balkanski, R. Wallis, and E. Haro, Anharmonic effects in light scattering due to optical phonons in silicon, \href{https://doi.org/10.1103/PhysRevB.28.1928}{\textcolor{blue}{Phys. Rev. B \textbf{28}, 1928 (1983)}}. 

\bibitem{JMenendez1984}
J. Menendez and M. Cardona, Temperature dependence of the first-order Raman scattering by phonons in Si, Ge, and $\alpha-$Sn: Anharmonic effects, \href{https://doi.org/10.1103/PhysRevB.29.2051}{\textcolor{blue}{Phys. Rev. B \textbf{29}, 2051 (1984)}}. 

\bibitem{JHasaien2025}
J. Hasaien, Y. Wu, M. Shi, Y. Zhai, Q. Wu, Z. Liu, Y. Zhou, X. Chen, and J. Zhao, Emergent quantum state unveiled by ultrafast collective dynamics in 1$T$-TaS$_2$, \href{https://doi.org/10.1073/pnas.2406464122}{\textcolor{blue}{Proc. Natl. Acad. Sci. USA  \textbf{122}, e2406464122 (2025)}}. 

\bibitem{BLoret2019}
B. Loret, N. Auvray, Y. Gallais, M. Cazayous, A. Forget, D. Colson, M. H. Julien, I. Paul, M. Civelli, A. Sacuto, Intimate link between charge density wave, pseudogap and superconducting energy scales in cuprates, \href{https://www.nature.com/articles/s41567-019-0509-5}{\textcolor{blue}{Nature Physics \textbf{15}, 771 (2019)}}.

\bibitem{DOh2021}
D. Oh, D. Song, Y. Kim, S. Miyasaka, S. Tajima, J. M. Bok, Y. Bang, S. R. Park, and C. Kim, B$_{1g}$-Phonon Anomaly Driven by Fermi Surface Instability at Intermediate Temperature in YBa$_2$Cu$_3$O$_{7{-\delta}}$, \href{https://doi.org/10.1103/PhysRevLett.127.277001}{\textcolor{blue}{Phys. Rev. Lett. \textbf{127}, 277001 (2021)}}.

\bibitem{HYHuang2021}
H. Y. Huang, A. Singh, C. Y. Mou, S. Johnston, A. F. Kemper, J. van den Brink $et$ $al$., Quantum Fluctuations of Charge Order Induce Phonon Softening in a Superconducting Cuprate, \href{https://doi.org/10.1103/PhysRevX.11.041038}{\textcolor{blue}{Phys. Rev. X \textbf{11}, 041038 (2021)}}.

\bibitem{XFTang2022}
X. Tang, S. Zhu, H. Liu, C. Zhang, Q. Wu, Z. Liu, J. Song, X. Guo, Y. Wang, H. Ma $et$ $al$., Growth, characterization, and Raman spectra of the 1T phases of TiTe$_2$, TiSe$_2$, and TiS$_2$, \href{https://iopscience.iop.org/article/10.1088/1674-1056/ac306a}{\textcolor{blue}{Chin. Phys. B \textbf{31}, 037103 (2022)}}.

\bibitem{SDuan2021}
S. Duan, Y. Cheng, W. Xia, Y. Yang, C. Xu, F. Qi, C. Huang, T. Tang, Y. Guo, W. Luo, $et$ $al$., Optical manipulation of electronic dimensionality in a quantum material, \href{https://www.nature.com/articles/s41586-021-03643-8}{\textcolor{blue}{Nature \textbf{595}, 239 (2021)}}.

\bibitem{HLiu2025}
H. Liu, C. Zhang, Q. Wu, Y. Jin, Z. Zhu, J. Song, S. Cui, Z. Sun, H. Wang, B. Chen $et$ $al$., Ultrafast photoinduced phase transition in the antiferromagnetic Dirac semimetal EuAgAs, \href{https://doi.org/10.1103/PhysRevB.111.L121113}{\textcolor{blue}{Phys. Rev. B \textbf{111}, L121113 (2025)}}.

\bibitem{YHMeng2024}
Y. Meng, Y. Yang, H. Sun, S. Zhang, J. Luo, L. Chen, X. Ma, M. Wang, F. Hong, X. Wang $et$ $al$., Density-wave-like gap evolution in La$_3$Ni$_2$O$_7$ under high pressure revealed by ultrafast optical spectroscopy, \href{https://www.nature.com/articles/s41467-024-54518-1}{\textcolor{blue}{Nat. Commun. \textbf{15}, 10408 (2024)}}.

\bibitem{Khasanov2025}
R. Khasanov, T. Hicken, I. Plokhikh, V. Sazgari, L. Keller $et$ $al$., Pressure and oxygen-isotope substitution on density-wave transitions in La$_4$Ni$_3$O$_{10}$, \href{https://doi.org/10.1103/nrqn-m22c}{\textcolor{blue}{Phys. Rev. Research \textbf{8}, 013249 (2026)}}.

\end{thebibliography}
\end{document}